\documentclass[aps,prl,reprint,superscriptaddress,nofootinbib]{revtex4-1}

\usepackage{graphicx}

\usepackage{amsmath,amssymb,mathrsfs}

\usepackage[bookmarks=false]{hyperref} %make sure it comes last of your loaded packages
\hypersetup{pdfstartview=FitH,pdfhighlight=/O,colorlinks=false}

\newcommand{\I}{\mathrm{i}}

\begin{document}

\title{Horizon energy as the boost boundary term in general relativity and loop gravity}
\author{Eugenio~Bianchi} 
\email{ebianchi@perimeterinstitute.ca}
\affiliation{Perimeter Institute for Theoretical Physics,
31 Caroline St.N., Waterloo ON, N2J 2Y5, Canada}
\author{Wolfgang~M.~Wieland}
\affiliation{Centre de Physique Th\`eorique, Campus de Luminy, Case 907,
   13288 Marseille, France}
\thanks{Unit\`e Mixte de Recherche (UMR 7332) du CNRS et de l'Universit\`e 
d'Aix-Marseille et de l'Univ Sud Toulon Var. Unit\`e affili\`ee \`a la
FRUMAM.}
\date{May 24, 2012}
%\date{\today}

\begin{abstract}
We show that the near-horizon energy introduced by Frodden, Ghosh and Perez arises from the action for general relativity as a horizon boundary term. Spin foam variables are used in the analysis. The result provides 
a derivation of the horizon boost Hamiltonian introduced by one of us to define the dynamics of the horizon degrees of freedom, and shows that loop gravity provides a realization of the horizon Schr\"odinger equation proposed by Carlip and Teitelboim. 
\end{abstract}

%\begin{flushleft}
%PACS: 04.60.Pp
%\end{flushleft}

\maketitle

In general relativity, the energy of an isolated gravitating system is coded into a boundary term at spatial infinity -- the ADM mass \cite{Arnowitt:1962hi}. In presence of a black hole, an inner boundary on the horizon can be introduced. In the case of a non-degenerate Killing horizon, Carlip and Teitelboim showed that the corresponding boundary term in the Hamiltonian generates boosts of the foliation near the horizon \cite{Carlip:1993sa}. This result was used by Massar and Parentani \cite{Massar:1999wg} and by Jacobson and Parentani \cite{Jacobson:2003wv} to study the thermalization of a stationary system close to the horizon including the back-reaction of the black hole. 

In this paper we consider the action for gravity used in spin foams \cite{Rovelli:2004tv}, the Einstein-Cartan action with Holst term in $BF$ variables, and show that there is a \emph{corner} term in the action associated to the horizon boundary. We prove the following three results: (i) the corner boundary term reproduces the Carlip-Teitelboim horizon Hamiltonian \cite{Carlip:1993sa}, (ii) the horizon Hamiltonian coincides with the horizon energy studied by Frodden, Ghosh and Perez \cite{Frodden:2011eb} when the proper time of a near-horizon stationary observer is used, (iii) in $BF$ variables, the horizon Hamiltonian is the function of the $B$-field that corresponds to the internal \emph{boost generator} upon spin-foam quantization. The internal boost Hamiltonian is exactly the operator used in \cite{Bianchi:2012fk} to code the dynamics of the quantum horizon in loop gravity. The result strengthens the loop gravity derivation of the thermodynamic entropy of non-extremal black holes. We note also that the horizon boundary term plays a key role in Smolin's analysis of the equation of state of spin foams where it is derived using similar methods \cite{LS-draft}.\\

In the spin foam approach to quantum gravity, the classical action considered is a functional of a Lorentz connection $\omega^{IJ}$ and a two-form $B^{IJ}$ given by the integral $I_4$ over a $4$-manifold $M_4$
\begin{equation}
I_4=\frac{1}{2}\int_{M_4}\big(\frac{1}{2}\epsilon_{IJKL}B^{KL}+\frac{1}{\gamma}B_{IJ}\big)\wedge F^{IJ}(\omega)\;,
\end{equation}
where $\gamma$ is the Immirzi parameter. At the classical level, this action coincides with the Einstein-Cartan action with a Holst term when the constraint $B^{IJ}=\frac{1}{8\pi G}e^I\wedge e^J$ is imposed. In spin foams, this constraint is imposed at the quantum level.

\begin{figure}[t]
\includegraphics[width=.3\textwidth]{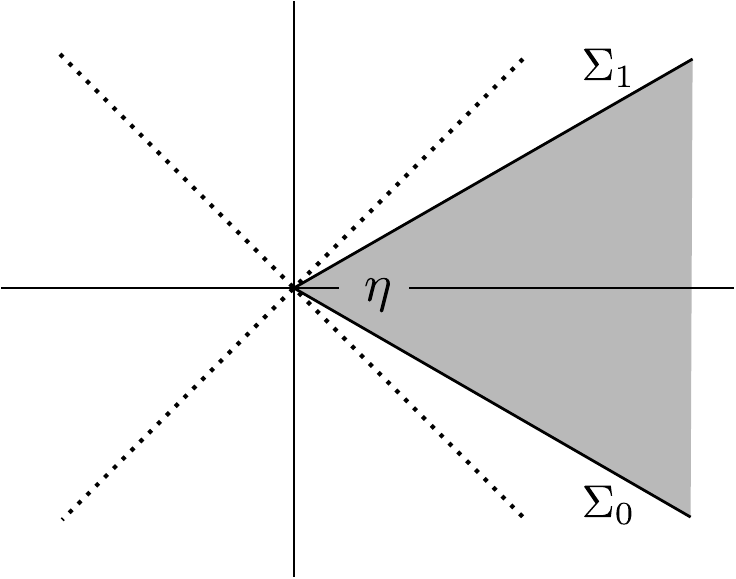}
\caption{The spatial sections $\Sigma_0$ and $\Sigma_1$ intersect on the horizon and bound a wedge with hyperbolic opening angle $\eta$.}
\label{fig:wedge}
\end{figure}

In presence of boundaries the bulk action $I_4$ has to be supplemented with boundary terms $I_3$ to ensure that classical solutions arise from a variational principle. Moreover, if the boundary is not smooth, corner terms $I_2$ are also present and the action $S$ consists of three contributions
\begin{equation}
S=I_4+I_3+I_2\;,
\end{equation}
as shown by Hayward, Hawking and Hunter \cite{Hayw:1993}. To describe an asymptotically-flat space with an inner boundary on a black-hole horizon we consider the region $M_4$ shown in figure (\ref{fig:wedge}). We call $\Sigma_0$ and $\Sigma_1$ the initial and the final spatial section and $\mathcal{T}$ the boundary at spatial infinity, so that $M_3=\partial M_4=\Sigma_0\cup\Sigma_1\cup\mathcal{T}$.  The region we consider is foliated by spatial sections $\Sigma$ that end all on a surface $M_2=\Sigma_0\cap\Sigma_1$ that correspond to the horizon. For a non-degenerate Killing horizon, sections $\Sigma$ of constant Killing time near the horizon can be used. Now we discuss boundary conditions on the horizon.

On the spatial sections $\Sigma$ we introduce a vector-valued $0$-form $n^{I}(x)$ that plays the role of  \emph{internal} time-like normal, $n^I n_I=-1$. The sections $\Sigma_0$ and $\Sigma_1$ intersect at the corner $M_2$ and induce on it the normals $n^I$ and $n'^I$ respectively. On $M_2$ we introduce also the vector-valued $0$-forms $z^{I}$ and $z'^{I}$ that play the role of  \emph{internal} space-like normals to $n^I$ and $n'^I$ respectively. We impose the following boundary conditions on our fields: on $\Sigma_0$ and $\Sigma_1$ we hold fixed the internal space-like components of the pullback of the connection, i.e. $\phi^*({h^I}_K{h^J}_L\delta \omega^{KL})=0$ where ${h^I}_J={\delta^I}_J+n^I n_J$. On the horizon $M_2$, we require $\delta(n_I z^I)=0$. When we discuss the tetrad $e^I$, we assume also that on the horizon $e^I n_I=e^I z_I=0$. Notice that we are not fixing the area of the horizon.

The boundary term $I_3$ allows to hold fixed the connection on the boundary in the variation of the  bulk term $I_4$. It is the familiar Hawking-York integral of the extrinsic curvature of $M_3$. In our variables it is given by
\begin{equation}
I_3=-\int_{M_3}\big(\frac{1}{2}\epsilon_{IJKL}B^{KL}+\frac{1}{\gamma}B_{IJ}\big)\wedge n^I D n^J\;,
\end{equation}
The role of $n^{I}$ is to hold fixed the space-like components of $\omega^{IJ}$ on $M_3$ in a covariant way \cite{Obuk:1987}. A gauge-fixed version of this expression can be found in \cite{Baez:1995jn}.

The corner term $I_2$ is less familiar. It is present whenever the variation of the normal to the boundary $M_3$ is not smooth. An example is the case of a cylindrical region in a Lorentzian spacetime, where the normal goes from space-like to time-like \cite{Hayw:1993}. We argue that the presence of an inner boundary on a black-hole horizon requires also a corner term. The region we consider is a wedge with the corner on the horizon as shown in figure (\ref{fig:wedge}): the evolution is frozen on $M_2$.  The corner term $I_2$ is given by
\begin{equation}
I_2=-\int_{M_2}\big(\frac{1}{2}\epsilon_{IJKL}B^{KL}+\frac{1}{\gamma}B_{IJ}\big) n^I z^J\;\,\eta\;,
\end{equation}
where $n^I$ and $z^I$ are the internal time-like normal and the internal space-like normal induced by the initial spatial section $\Sigma_0$ on $M_2$. The quantity $\eta$ is a hyperbolic angle defined by
\begin{equation}
\cosh \eta\;=-\;n_I\;n'^I\;,
\end{equation}
with $n'^I$ the time-like normal induced by the final spatial section $\Sigma_1$ of the wedge. The boundary term $I_2$ allows to hold $n_Iz^I=0$ fixed in the variation of the action with respect to $n^I$. We refer to \cite{BW:defects} for a detailed derivation.\\

Now we come to our three results. Imposing boundary conditions $e^I\, z_I=0$ on the tetrad on the horizon, we recognize that, for $B^{IJ}=\frac{1}{8\pi G}e^I\wedge e^J$, the corner term $I_2$ is the integral of the area density on the horizon. Calling $A$ the area of $M_2$ with the orientation induced by $\Sigma_0$,
\begin{equation}
A=\int_{M_2} \frac{1}{2}\epsilon_{IJKL}(e^K\wedge e^L)\;n^I z^J\;,
\end{equation} 
we find 
\begin{equation}
I_2=-\frac{A}{8\pi G}\; \eta\;.
\end{equation}
This is exactly the expression derived by Carlip and Teitelboim in \cite{Carlip:1993sa} using the metric formalism: the boundary term $I_2$ is the product of a horizon Hamiltonian times an evolution parameter $\eta$. The parameter $\eta$ codes the boost from the initial spatial section $\Sigma_0$ to the final spatial section $\Sigma_1$.\\

Being a hyperbolic angle, the parameter $\eta$ is dimensionless. Let us consider a foliation of $M_4$ in spatial sections $\Sigma$ that near the horizon have constant $\eta$. The proper time $t$ of a stationary observer at small distance $\ell$ from the horizon is $t=\eta\,\ell$. As a result, if we use $t$ as time parameter we find that the boundary term assumes the form
\begin{equation}
I_2=-E\;t\;,
\end{equation}
where the quantity $E$ has dimensions of energy and is given by
\begin{equation}
E=\frac{A}{8\pi G}\,\ell^{-1}\;.\label{eq:FGP}
\end{equation}
This expression coincides exactly with the near-horizon energy derived by Frodden, Ghosh and Perez considering the back-reaction of the black-hole geometry when a near-horizon stationary observer lets some matter fall through the horizon \cite{Frodden:2011eb}.\\

Now we consider the quantum theory. From the bulk term $I_4$ of the action it is easy to see that the momentum $\Pi_{IJ}$ conjugated to the Lorentz connection $\omega^{IJ}$ is given by
\begin{equation}
\Pi_{IJ}= \frac{1}{2}\epsilon_{IJKL}B^{KL}+\frac{1}{\gamma}B_{IJ}\;.
\end{equation}
In the quantum theory, the associated operator acts on states $\Psi[\omega]$ as a functional derivative
\begin{equation}
\hat{\Pi}_{IJ}=-\I\hbar\frac{\delta}{\delta \omega^{IJ}}\;.
\end{equation}
In the spin foam approach to quantum gravity where holonomies of the Lorentz connection are used, the operator $\Pi_{IJ}$ is simply the generator of Lorentz transformations in a unitary representation of $SL(2,\mathbb{C})$ belonging to the principal series $\mathcal{V}^{(p,k)}$. It is useful to introduce the generators of boosts and rotations that preserve the time-like vector $n^I$,
\begin{equation}
K^f_I=\frac{1}{\hbar}n^K\,\hat{\Pi}^f_{KI}\quad,\quad\;  L^f_I=\frac{1}{2\hbar}\epsilon_{KIJL}\,n^K\,\hat{\Pi}_f^{JL}\;.\label{eq:boost}
\end{equation}
As these operators act on holonomies, they are labeled not by points $x$ but by faces $f$ of a 2-complex in space-time. The constraint $B^{IJ}=\frac{1}{8\pi G}e^I\wedge e^J$ that reduces the action $I_4$ to gravity is imposed at the quantum level selecting $\gamma$-simple representations of $SL(2,\mathbb{C})$. Such representations solve weakly the constraint $K_f^I\,\approx\,\gamma\, L_f^I$ on each face \cite{Rovelli:2004tv,Engle:2007uq}.\\

In terms of the momentum $\Pi^{IJ}$, the corner term is simply given by $I_2 = -H \eta$ where
\begin{equation}
H=\int_{M_2}\Pi_{IJ}\; n^I z^J\;.
\end{equation}
Its spin foam quantization is immediate: the surface $M_2$ is tessellated in faces $f$ and using (\ref{eq:boost}) the two-form $\Pi_{IJ}\; n^I z^J$ is recognized to be the component of the boost $\hat{K}^f_z=K^f_I z^I$ in the direction $z^I$ orthogonal to $M_2$,
\begin{equation}
\hat{H}=\sum_f \hbar\,K_z^f\;.
\end{equation}
This operator generates boosts of each facet $f$ with parameter $\eta$. The Hamiltonian describing the dynamics in proper time $t$ is given by $\hat{E}=\hat{H} \ell^{-1}$. This is exactly the boost Hamiltonian used in \cite{Bianchi:2012fk} to define the dynamics of the quantum horizon. The thermality of the quantum horizon arises from the properties of the boost Hamiltonian $\hat{H}$. Moreover, this operator is consistent with the classical expression (\ref{eq:FGP}) that is reproduced as the expectation value on an eigenstate of the area $\hat{A}=8\pi G\hbar \gamma \sum_f|L_f^I|$ of the horizon, 
\begin{equation}
\langle{\textstyle \sum_f \hbar\,K_z^f \,\ell^{-1}}\rangle=\frac{A}{8\pi G}\ell^{-1}\;,\label{eq:<E>}
\end{equation}
as shown in \cite{Bianchi:2012fk} .\\

In \cite{Carlip:1993sa}, Carlip and Teitelboim proposed that to the horizon Hamiltonian corresponds a horizon Schr\"odinger equation
\begin{equation}
\I\hbar\frac{\partial}{\partial \eta}\psi\;=\;\hat{H}\,\psi
\end{equation}
that generates evolution of states in boost time. Loop gravity provides a specific realization of this proposal. The new feature appearing in loop gravity is that the horizon Hamiltonian is not simply proportional to the area operator as conjectured in \cite{Carlip:1993sa,Massar:1999wg,Jacobson:2003wv}, it is truly the horizon boost generator. In fact the area operator does not commute with the Hamiltonian as rotations don't commute with boosts. The relation between the area and the energy arises only as expectation values. The dispersion is controlled by a horizon energy-area uncertainty relation \cite{Bianchi:2012fk}.\\

The appearance of the proper distance $\ell$ in (\ref{eq:FGP}) and in (\ref{eq:<E>})  can lead to the impression that the horizon Hamiltonian is a property of some specific observer who is stationary near the horizon. In this paper we argued that the boost Hamiltonian is truly a property of gravity itself and we made this essential feature manifest showing that the Hamiltonian arises as a horizon boundary term in general relativity.

\vspace{2,5em}

\emph{Acknowledgements}. We would like to thank T.~Jacobson, C.~Rovelli and L.~Smolin for useful discussions. WW thanks Perimeter Institute for hospitality and support. Research at Perimeter Institute for Theoretical Physics is supported in part by the Government of Canada through NSERC and by the Province of Ontario through MRI.

%\bibliographystyle{utcaps}
%\bibliography{omnibib}

\end{document}